\documentclass[preprint]{vgtc}               

\usepackage{mathptmx}
\usepackage{graphicx}
\usepackage{times}

\onlineid{0}

\vgtccategory{Research}

\vgtcinsertpkg

\preprinttext{Presented as a Poster at IEEE VIS 2016 conference in Baltimore, MD, USA}

\title{ClassSPLOM - A Scatterplot Matrix \\to Visualize Separation of Multiclass Multidimensional Data}

\author{Michael Aupetit\thanks{e-mail: maupetit@hbku.edu.qa (Corresponding author)}\\ %
        \scriptsize Qatar Computing Research Institute, HBKU, Doha, Qatar %
\and Ahmed Ali\thanks{e-mail: amali@hbku.edu.qa}\\ %
     \scriptsize Qatar Computing Research Institute, HBKU, Doha, Qatar  
}

\abstract{In multiclass classification of multidimensional data, the user wants to build a model of the classes to predict the label of unseen data. The model is trained on the data and tested on unseen data with known labels to evaluate its quality. The results are visualized as a confusion matrix which shows how many data labels have been predicted correctly or confused with other classes. The multidimensional nature of the data prevents the direct visualization of the classes so we design ClassSPLOM to give more perceptual insights about the classification results. It uses the Scatterplot Matrix (SPLOM) metaphor to visualize a Linear Discriminant Analysis projection of the data for each pair of classes and a set of Receiving Operating Curves to evaluate their trustworthiness. We illustrate ClassSPLOM on a use case in Arabic dialects identification.

} 

\begin{document}


\firstsection{Introduction}

\maketitle


In a multiclass classification setting,  there are two main approaches to  model the classes aggregating the results of either one-vs-all or one-vs-one binary classifiers \cite{Bishop2006}. The user evaluates the quality of a multiclass classifier by visualizing the so-called \emph{confusion matrix}.

This matrix is a numerical table indicating the number of data within a given class that have been predicted to be of the same or of another class (Table \ref{tab_confusion}). Among other possible scores,   \emph{Precision} (PRC)  summarizes for each predicted class the proportion that comes from the corresponding true class, and \emph{Recall} (RCL) summarizes for each true class the proportion which is predicted correctly.

However due to the multidimensional nature of the data, the user lacks insights to explain the classification results especially to know how to improve the model when the results are bad (low PRC or RCL). Different metaphors have been proposed to display the confusion matrix but without providing new information. 

Here we propose a new metaphor which consists in computing linear classifiers for each pair of classes and visualize their result in a scatterplot matrix. This visualization shows how much two classes are linearly separable or overlap each other and the other classes. We complement each scatterplot with a Receiving Operating Curve (ROC) which allows  evaluating its trustworthiness.

\section{Related work}
 
Scatterplot Matrices (SPLOM) \cite{Munzner2014} align scatterplots relative to pairs of variables, with one variable per row or per column. This alignment eases the comparison between the plots. Usually, the up  or low triangular part of the matrix is filled with scatterplots while the other part is empty or filled with complementary information. In the case of multiclass data classification, the user needs to focus on the pairwise separation of the classes. So we propose to use the SPLOM metaphor but considering pairs of classes instead of pairs of variables. 
The confusion matrix can be visualized as a heatmap by coloring the cells according to their value. A stacked bar chart metaphor has also been proposed in \cite{Beauxis2015} which helps the user understand the confusion between desired and predicted labels. 

Our focus is slightly different here, we do not visualize the confusion matrix itself but visualize complementary projections of the original data to shed new light on the classification results. In \cite{Aupetit2014}, a matrix of class-color-coded histograms is proposed to visualize for each pair of classes the distribution of the data along a projection axis joining the center of gravity of both classes. The histogram quantifies the data into bins along the x-axis and exploits the y-axis dimension to inform about the count in each bin. We propose to improve this approach by replacing the histograms by  \emph{LDA scatterplots} which provide more information within the same space, and to use the second half of the matrix to display \emph{ROC-AUCBA plots} showing their trustworthiness.

\section{ClassSPLOM}

\subsection{Objective}

The objective is to give visual insight about the separation or overlapping of each pair of classes in the multidimensional space. There is no unique way to define class separation as any classifier gives its own definition. We propose to consider the linear classifier defined by Fisher Linear Discriminant Analysis (LDA)\cite{FisherLDA} as a basis for comparison to the possibly non-linear classifier used to get the confusion matrix. 
If two classes appear separated along the LDA axis, so are they in the multidimensional space and the confusion matrix should show low confusion between them. If they are mixed along this axis, then they are still possibly separated in a nonlinear way LDA cannot show.

\subsection{Design}

\begin{figure}[htb]
  \centering
  \includegraphics[width=2.5in]{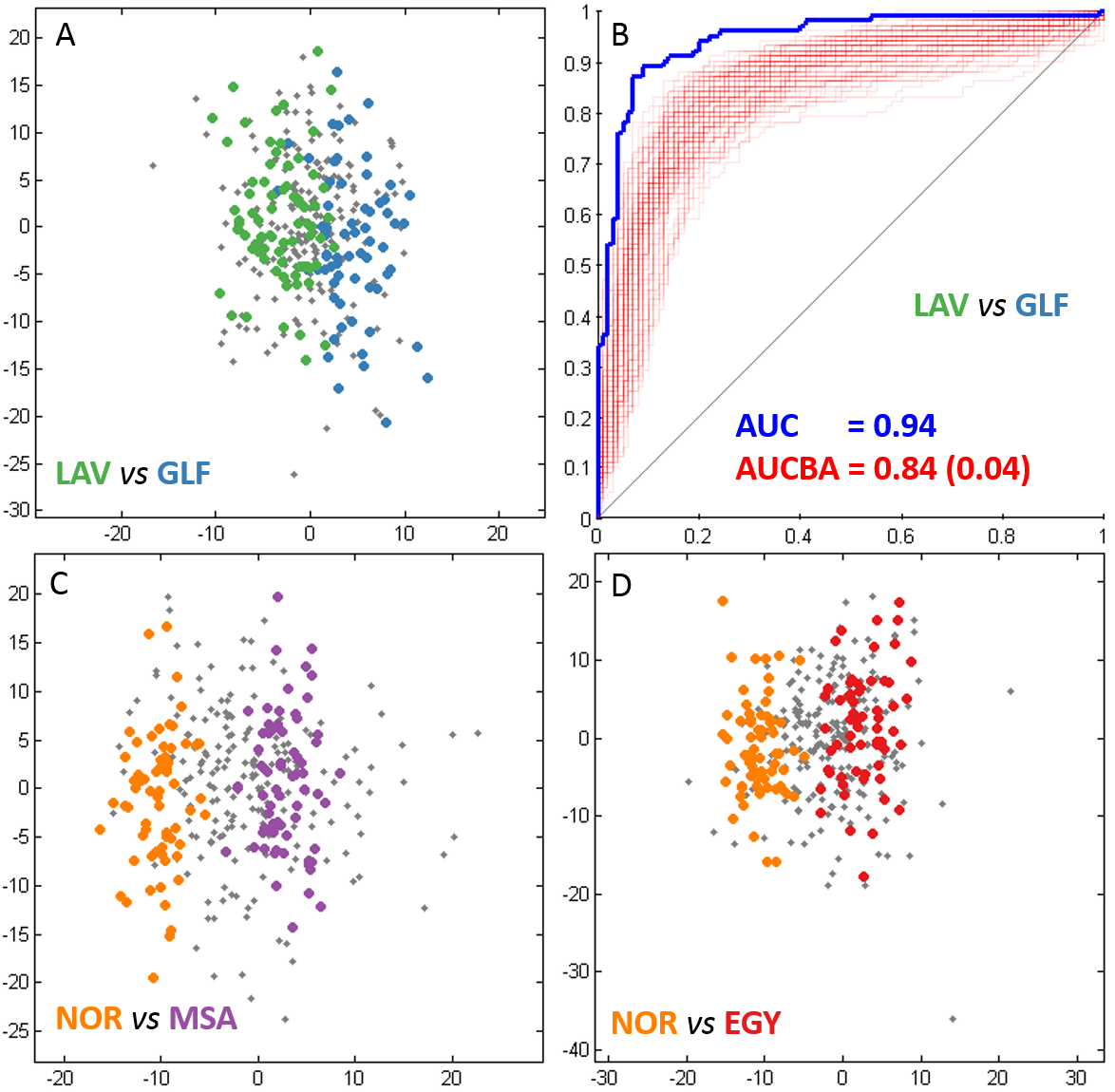}
  \caption{LDA scatterplots (A,C,D) and ROC-AUCBA plot (B).}
  \label{fig_Detail}
\end{figure}

\vspace{-0.3cm}
\noindent\paragraph{LDA scatterplot:} (Fig.\ref{fig_Detail}ACD) For a pair of classes $(c_1,c_2)$, the x-axis is the linear discriminant axis computed by LDA between $c_1$ and $c_2$. The y-axis is the LDA axis of the data projected in the $D-1$ subspace orthogonal to the first LDA axis, and computing the LDA between classes $c_{12}=c_1\cup c_2$ versus the rest. These two axes are orthogonal in the data space and maximize the ratio of between-class variance over within-class variance of the projected data. In case there are only two classes, the second LDA can be replaced by the first principal component in the subspace orthogonal to the first axis. Data of both classes are color-coded using a qualitative color scale. Data from any other class are displayed in background with a grey color. Aspect ratio is 1 and unit vectors are of equal length.

\vspace{-0.2cm}
\noindent\paragraph{ROC-AUCBA plot:} (Fig. \ref{fig_Detail}B) The Receiving Operating Curve (ROC) \cite{fawcett2006} is computed by thresholding the x-axis of the corresponding scatterplot. However each scatterplot is a \emph{descriptive} view of the data which gives no insight about the \emph{predictive} power of the x-axis. It is biased toward the data sample (overfitting). We need to show how much this graphical representation is predictive of the class separation we would observe if we were to display new data coming from the same statistical distribution of the classes. For this sake we generate 100 bootstrap samples and for each we compute the LDA projection, its ROC and Area Under the ROC (AUC), a standard quality measure of binary classifiers ranging form 0.5-to-1 for random-to-perfect classification. We average the AUC over the 100 bootstrap samples to get the AUC Bootstrap Average (AUCBA) and its standard deviation. The AUCBA is indicative of the predictive power of the  displayed descriptive LDA scatterplot. We render the bootstrap ROCs as red curves with transparency set to 0.1 and the observed ROC as an opaque blue line.  

\vspace{-0.2cm}
\noindent\paragraph{ClassSPLOM:} (Fig. \ref{fig_ClassSPLOM}) To generate the ClassSPLOM metaphor, we draw a large class-color-coded disc into each cell of the matrix diagonal, showing the name of the class. The diagonal elements serve as an in-situ legend indicating the color-code of the classes displayed in the LDA scatterplots, and the spatial relation between rows, columns and classes. The Gestalt law of symmetry is used with respect to the matrix diagonal to support visual association of a scatterplot to its corresponding ROC-AUCBA plot.

\vspace{-0.2cm}
\subsection{Use case}

We use the ClassSPLOM metaphor to analyze the results of an Arabic dialect identification task \cite{Interspeech2016}. 
The Natural Language Processing (NLP) community has aggregated
dialectal Arabic into five regional language groups:
Egyptian (EGY), North African or Maghrebi (NOR), Gulf or
Arabian Peninsula (GLF), Levantine (LAV), and Modern Standard
Arabic (MSA). The user is an expert in linguistics and speech processing. He assumes that Arabic dialects are historically but not synchronically related, so if they are mutually unintelligible languages like English and Dutch, classes should appear separated from each other. 

\begin{figure}[htb]
  \centering
  \includegraphics[width=3in]{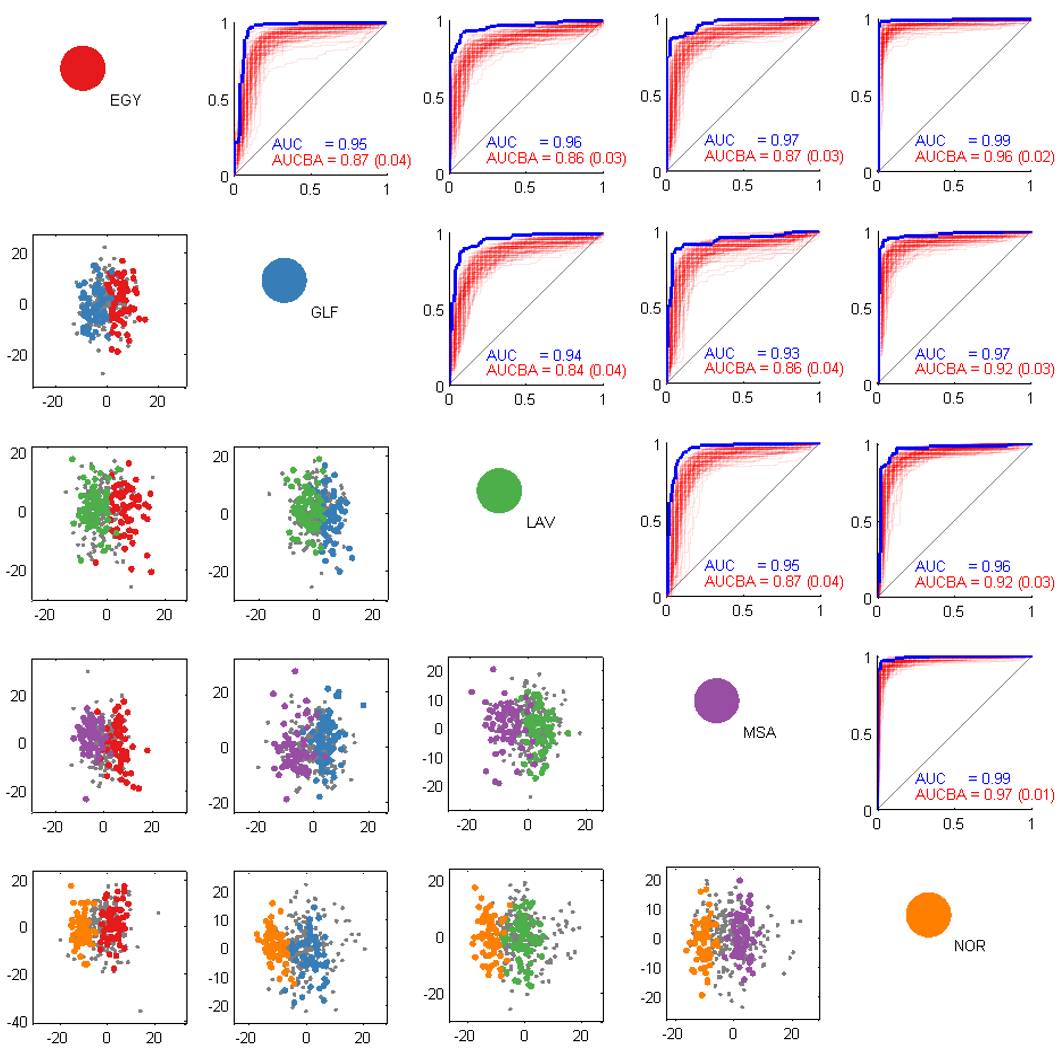}
  \caption{ClassSPLOM visualization of multiclass data.}
  \label{fig_ClassSPLOM}
\end{figure}

Data come as $8802$ speech utterances from the $5$ dialects encoded in a $569610$ feature space  \cite{Interspeech2016}. We subsampled them to $100$ points per class and projected them in $100$ dimensions using Principal Component Analysis. The LDA projection is performed in this reduced space.
The confusion matrix obtained using a nonlinear classifier \cite{Interspeech2016} is shown in table \ref{tab_confusion} and ClassSPLOM in the figure \ref{fig_ClassSPLOM}.

\begin{table}
  \caption{Confusion matrix of the Arabic dialect identification task. PRC and RCL scores read $77\%$ of predicted MSA are actually MSA, and $78.1\%$ of real MSA are predicted MSA respectively.}
  \label{tab_confusion}
  \scriptsize
  \begin{center}
    \begin{tabular}{cccccccc}
    \hline
       & EGY & GLF & LAV & MSA & NOR & \#True  & RCL\%\\
    \hline
      EGY & \textbf{221} &  15 & 57 & 13 & 9 & \textbf{315} & 70.2\\
      GLF & 45 & \textbf{121} & 82 & 12 & 5 & \textbf{265} & 45.7\\
      LAV & 74 & 43 & \textbf{199} &  18 & 14 & \textbf{348} & 57.2\\
      MSA & 19 & 17 & 20 & \textbf{218} & 5 & \textbf{279} & 78.1\\
      NOR & 80 & 21 & 66 & 22 & \textbf{166} & \textbf{355} & 46.8\\
      \#Pred & \textbf{439} & \textbf{217} & \textbf{424} &\textbf{283} &\textbf{199} & &\\
      PRC\% & 50.3 & 55.8 & 46.9 & 77 & 83.4 & &\\
    \hline
    \end{tabular}
  \end{center}
\end{table}

ClassSPLOM shows that GLF and LAV are the less linearly separable classes (Fig. \ref{fig_ClassSPLOM}), which gives perceptual confirmation of the high confusion values in the confusion matrix. The $0.94$ AUC score and blue ROC compared to the $0.84$ AUCBA score and distribution of red bootstrap ROCs (Fig. \ref{fig_Detail}B) tells that the visible separation of the classes (Fig. \ref{fig_Detail}A) is actually optimistic. Conversely, NOR and EGY and NOR and MSA classes are clearly linearly separated  (Fig. \ref{fig_Detail}CD) and the visible linear separation is trustworthy according to the AUCBA score, which perceptually confirms the low confusion between these pairs of classes in the confusion matrix. 
These findings support the user expectations.

\section{Discussion}

ClassSPLOM provides a perceptual overview of the classes' relations in terms of linear separation, that is more difficult to catch reading the symbolic confusion matrix. However the overview seems to show good separation of each class with respect to each other, despite the low Precision and Recall scores obtained using a supposed-to-be better nonlinear classifier. The reason is that separated colored points in a scatterplot can overlap in another one. We plan to help the user detect this issue by enabling color on demand of the background grey points, and by implementing brush and link across the views. We also want to improve this approach by distinguishing between training and test sets used to get the confusion matrix, and by linking the scatterplots to its cells.


\bibliographystyle{abbrv}
\bibliography{ClassSPLOM}
\end{document}